\documentclass[10pt]{iopart}
\usepackage{geometry}                
\usepackage[parfill]{parskip}    
\usepackage{graphicx}
\usepackage{amssymb}
\usepackage{epsfig}
\usepackage{epsf}
\usepackage{epstopdf}
\usepackage[hypertex]{hyperref}

\begin{document}
\title{Diffusion of active tracers in fluctuating fields}

\author{David S. Dean and V. D\'emery}

\address{Laboratoire de Physique Th\'eorique -- IRSAMC,
Universit\'e de Toulouse, CNRS, 31062 Toulouse, France}

\pacs{}

\begin{abstract}
The problem of a particle diffusion in a fluctuating scalar field is studied. In contrast to most studies
of advection diffusion in random fields we analyze the case where the particle position is also 
coupled to the dynamics of the field. Physical realizations of this problem are numerous and range
from the diffusion of proteins in fluctuating membranes and the diffusion of localized  magnetic fields 
in spin systems. We present exact results for the diffusion constant of particles diffusing in dynamical Gaussian fields in the adiabatic limit where the field evolution is much faster than the particle diffusion.
In addition we compute the diffusion constant perturbatively, in the weak coupling limit where  the
interaction of the particle with the field is small, using a Kubo-type relation. Finally we construct a simple toy model which can be solved exactly.
\end{abstract} 

\maketitle

\section{Introduction}
The diffusion of passive particles in complex and random velocity fields has been extensively studied in
statistical mechanics and fluid mechanics. In most of these studies one is interested in the dispersion 
of tracer particles which are advected by the complex or random field but which do not affect the field
itself- it is for this reason they are called passive \cite{shr2000,fal2001}. These systems can be studied in the case of  incompressible velocity fields (relevant for problems of turbulent dispersion in fluid mechanics) and the case where the velocity field is derived from a gradient (relevant to statistical mechanics and dynamical transitions related to the glass transition). As well as diffusion in time dependent fields the problem of diffusion in quenched random fields has also
been extensively studied \cite{revq}.  Assuming that the advecting velocity field has zero mean, the 
passive tracer particle  may diffuse normally as  $\langle {\bf x}^2(t)\rangle \sim t$ but in
certain circumstances  the particle diffuses  anormally $\langle {\bf x}^2(t)\rangle \sim t^{2\nu}$ with $\nu \neq1/2$. In the case where $\nu<1/2$ the diffusion is called subdiffusive and when $\nu>1/2$ the diffusion is called superdiffusive \cite{bou1990}. Other interesting phenomena arise when one considers the diffusion of an ensemble of non-interacting (among themselves) tracers. Depending on the 
statistics of the advecting field clustering phenomena may arise \cite{fal2001}.  

In this paper we will consider a problem where the tracer particle's position is coupled to the 
evolution of a scalar field. This means that the particle is advected by the field but also the dynamics of the field is affected by the particle position. We take the dynamics of the field to be over damped 
stochastic dynamics. An interesting question arises as to how the diffusion of the 
particle depends on the fields's dynamics, for instance one could have the same Hamiltonian for the system but in one case the field may evolve according to model A (non-conserved order parameter)
dynamics and in the other by model B (conserved order parameter dynamics) or indeed Brownian 
hydrodynamics \cite{cha2000}. We will first concentrate our study on the case where the  field dynamics is much more rapid than the local diffusion of the tracer particle. In this limit we  will show that the diffusion of the particle is always slowed down by coupling the field. This is in contrast to the case where the effect of the tracer on the particle is ignored and where in this limit it can be shown that the diffusion of the particle is speeded up. We then consider a perturbative calculation of the effective diffusion 
constant where the coupling between the field and particle position is weak. This computation 
is based on a Kubo-like relation for the effective diffusion constant, in addition this relation shows 
explicitly that the diffusion constant of the particle is reduced when its position is coupled to
the fluctuating field. 

A concrete example of this problem is the diffusion of a protein on a membrane. The protein's
position can be coupled to the height fluctuations of the membrane, for instance by tending to 
impose a local mean curvature \cite{gou1993}. Alternatively the protein's position could be coupled to the  local composition of the membrane, this could be because the protein has an affinity for a 
particular lipid type in a multicomponent lipid membrane or because it imposes a local tilt on the
lipid hydrocarbon tails \cite{sac1995}. These two types of couplings lead to membrane mediated
interactions between proteins but they will also modify the dynamics of a protein in the membrane.
Indeed the question of what determines a proteins diffusion constant in a lipid bilayer is biologically very important. The problem was first addressed by Saffmann and Delbr\"uck \cite{saff1975} based on a two dimensional fluid model. However experimental studies \cite{gam2006} suggest that this simple fluid model may not explain the experimental data on protein and peptide diffusion constants as a function of their size.  A number of studies have subsequently shown that protein coupling to membrane composition and height can substantially modify the protein's diffusion constant \cite{rei2005,naji2007,lei2008,naji2009,dem2010a,dem2010b,lei2010}. We should note that height fluctuations can modify the effective observed diffusion constant of a protein in a membrane even when there is no coupling between the fluctuations and protein position, this is because the protein diffuses in the plane of the membrane but the diffusion is observed on the projected area, this leads to an apparent reduction  of the diffusion constant, this is a geometrically induced effect \cite{lin2004,naji2007a}.

\section{The model}

Consider the dynamic of a Langevin particle whose position is denoted by ${\bf x}(t)$ diffusing with a linear coupling to a fluctuating free field. The overall Hamiltonian for the system is
\begin{equation}
H= {1\over 2}\int \phi({\bf x})\Delta\phi({\bf x}) d{\bf x} - hK\phi({\bf x}(t))
\end{equation}
where we will take $\Delta$ and $K$ to be self adjoint operators. The above Hamiltonian applies
to many systems. For example if we take $\Delta = -\nabla^2 +m^2$ and $K= 1$, this is a model for
a point magnetic field of magnitude $h$ diffusing in a Gaussian ferromagnetic model. If
$\Delta = \kappa\nabla^4 -\sigma\nabla^2$ and $K=-\nabla^2$, the Hamiltonian is the Helfrich one
for the height fluctuations of a lipid membrane where $\phi$ represents the height and the choice of
$K$ is due to the fact the particle is coupled to the local membrane curvature \cite{hel1973}. Here we
are interested in the diffusion of the particle in the field. However to study the dynamics of diffusion we must also define the dynamics of the field. Here we will take for the field dynamics the general
dissipative dynamics form
\begin{equation}
{\partial\phi({\bf x})\over \partial t} = -\kappa_\phi R{\delta H\over \delta\phi({\bf x})} + \sqrt{\kappa_\phi}\xi({\bf x},t)
\end{equation}
and where $R$ is a self adjoint dynamical operator and 
$\xi$ is a Gaussian noise of zero mean which is uncorrelated in time. For instance, $R=\delta({\bf x}-{\bf x}')$ corresponds to model A conserved dynamics and $R=-\nabla^2\delta({\bf x}-{\bf x}')$ corresponds to model B dynamics \cite{cha2000}. If one is considering the case where $\phi$ represents the height fluctuations of 
a membrane then using Brownian hydrodynamics the Fourier transform of $R$, $\tilde R$ is given by
$\tilde R({\bf k})= 1/4\eta |{\bf k}|$, where $\eta$ is the viscosity of the solvent surrounding the membrane
\cite{lin2004}.  The field dynamics is taken so as to respect detailed balance so that the Gibbs-Boltzmann  distribution is found for the equilibrium measure of the field and the particle position. This means that

\begin{equation}
\langle \xi({\bf x},t)\xi({\bf x}',t')\rangle  =2T R({\bf x}-{\bf x}')\delta(t-t'),
\end{equation}
where $T$ is the temperature of the system. The dynamics of the particle is given by
\begin{equation}
{\partial x_i(t)\over \partial t} = -\kappa {\partial H\over \partial x_i} + \sqrt{\kappa}\eta_i(t),
\end{equation}
where the noise terms is Gaussian  noise with zero mean zero and correlation function
\begin{equation}
\langle \eta_i(t)\eta_j(t')\rangle = 2T \delta_{ij}\delta(t-t').
\end{equation}

The coefficients $\kappa$ and $\kappa_\phi$ can be used to set the relative time scale between the dynamics of the field fluctuations and that of the tracer movement. In the absence of a coupling between the field and the particle, the particle diffuses normally and within the notation set up here the mean 
squared displacement at large times behaves as
\begin{equation}
\langle {\bf x}^2(t)\rangle \sim 2dT\kappa t= 2d Dt,\label{dodef}
\end{equation}
where $d$ is the spatial dimension and $D = T\kappa$ is the bare diffusion constant. For the specific choice of Hamiltonians considered here we thus have the equations of motion

\begin{equation}
{\partial\phi({\bf x})\over \partial t} = -\kappa_\phi R\Delta \phi({\bf x})  
+ h \kappa_\phi RK(x-{\bf x}(t)) + \sqrt{\kappa_\phi}\xi({\bf x},t)  \label{phid}
\end{equation}
and
\begin{equation}
{\partial x_i(t)\over \partial t} = h\kappa\nabla K\phi({\bf x}(t)) + \sqrt{\kappa}\eta_i(t),\label{xd}
\end{equation}
where in this paper, for two operators $A$ and $B$ will will denote by $AB$ there composition as operators. In the limit where it is defined we will be interested in the effective diffusion constant 
for the tracer defined via
\begin{equation}
\langle {\bf x}^2(t)\rangle \sim 2dT\kappa_e t= 2d D_et,
\end{equation}
where $D_e$ is the effective late time diffusion constant. 

We note that, as mentioned above,  Eq. ({\ref{xd}) has been extensively studied in the case where the field $\phi$ evolves  independently of the particle position. This problem is referred to as the advection diffusion of a passive scalar (the concentration of the particle) in a fluctuating field $\phi$. This is obtained in the limit where one sets $h=0$ in Eq. (\ref{phid}) but keeps $h\neq 0$ in Eq. (\ref{xd}). It was suggested that this limit can be used to approximate the diffusion of the tracer particle in \cite{rei2005,lei2008}. In this case it is found that the effect of the field fluctuations can be to increase the diffusivity of the tracer particle with respect to that obtained when it is not coupled to the fluctuating field ($h=0$). However the numerical simulations of \cite{naji2009} where the effect of the particle position on the field is taken into account suggests that the diffusion is reduced with respect to the case $h=0$
and in \cite{lei2010} the authors of \cite{rei2005,lei2008} revisited the problem and 
numerically and analytically confirmed the findings of \cite{naji2009} . The adiabatic results  obtained here show that in this limit the diffusivity is always diminished with respect to the case $h=0$ and we argue that this is the limit where the diffusivity should be the most rapid. We also will show via a Kubo-type formula that active coupling to the fluctuating field should always reduce the
value of the diffusion constant.

 \section{The adiabatic limit}

We will now analyze the dynamics of this system in the limit $\kappa_\phi \gg \kappa$, i.e. where the field dynamics is much quicker than that of  the particle. The basic idea is that one can eliminate the 
field variable in a mathematically controlled manner to yield an effective diffusion equation for the particle where the field no longer appear implicitly. This sort of procedure can be carried out at the 
level of the Fokker Planck equation using projection operator methods \cite{risk1996}. However in the case here as we have a dynamical variable $\phi$ with an infinite number of degrees of freedom
we will use an alternative method based on direct analysis of the Langevin equations 
\cite{san1980,tou2009}.

We start by integrating the equation of motion of the field $\phi$ to obtain:
\begin{equation}
\phi({\bf x}) = \int_0^t ds\ \exp(-\kappa_\phi (t-s)R\Delta)
\left[h\kappa_\phi RK({\bf x}-{\bf x}(s)) +\sqrt{\kappa_\phi}\xi({\bf x},s)\right].
\end{equation}
Now if $\kappa_\phi$ is large and the operator $R\Delta$ is positive, the above integral is
dominated by the region where $s$ is close to $t$. We make the simple change of variables
$u=t-s$ in the above to find
\begin{equation}
\phi({\bf x}) = \int_0^t du\ \exp(-\kappa_\phi uR\Delta)
\left[h\kappa_\phi RK({\bf x}-{\bf x}(t-u)) +\sqrt{\kappa_\phi}\xi({\bf x},t-u)\right].
\end{equation}
A Taylor expansion  about $u=0$ now yields for large $t$, 
\begin{eqnarray}
\phi({\bf x}) &=& \int_0^t du\ \exp(-\kappa_\phi uR\Delta)
\left[h\kappa_\phi RK({\bf x}-{\bf x}(t) )+ u{dx_j(t)\over dt}\nabla_j RK({\bf x}-{\bf x}(t) )+\sqrt{\kappa_\phi}\xi({\bf x},t)\right] \nonumber \\ &+&O({1\over \kappa_\phi^{3\over 2}}) \nonumber \\
&=& h\Delta^{-1}K({\bf x}-{\bf x}(t)) +{h\over \kappa_\phi}{dx_j(t)\over dt}(R\Delta)^{-2}\nabla_jRK({\bf x}-{\bf x}(t) ) +\sqrt{{1\over \kappa_\phi}}(R\Delta)^{-1}\xi({\bf x},t) \nonumber \\\label{exp}
\end{eqnarray}
We must now compute $\nabla K\phi({\bf x}(t))$ from the above. We may write the first term of 
Eq. (\ref{exp}) using its Fourier representation as
\begin{equation}
hK\Delta^{-1}K({\bf x}-{\bf x}(t))  =  {h\over (2\pi)^d}\int d{\bf k} {\tilde K^2(k)\over \tilde \Delta(k)}\exp\left(i{\bf k}\cdot
({\bf x}-{\bf x}(t))\right)
\end{equation}
and the second term is given by
\begin{equation}
{dx_j(t)\over dt}K(R\Delta)^{-2}\nabla_jRK({\bf x}-{\bf x}(t) ) ={dx_j(t)\over dt}{h\over\kappa_\phi (2\pi)^d}\int d{\bf k}\  ik_j{\tilde K^2(k)\over \tilde\Delta(k)^2 \tilde R(k)} \exp\left(i{\bf k}\cdot
({\bf x}-{\bf x}(t))\right).
\end{equation}
The results now give that to the order of approximation in $1/\kappa_\phi$ used above we have
\begin{eqnarray}
\nabla_iK\phi({\bf x}(t))
&=& \nonumber  {h\over (2\pi)^d}\int d{\bf k}\ ik_i {\tilde K^2(k)\over \tilde \Delta(k)}-
{dx_j(t)\over dt}{h\over\kappa_\phi (2\pi)^d}\int d{\bf k}\  k_jk_i{\tilde K^2(k)\over \tilde\Delta(k)^2 \tilde R(k)}\nonumber \\
&+& \sqrt{{1\over \kappa_\phi}}\nabla_iK(R\Delta)^{-1}\xi({\bf x},t) 
\end{eqnarray}
The first term is zero by isotropy  and we can also write
\begin{equation}
\int d{\bf k}\  k_jk_i{\tilde K^2(k)\over \tilde\Delta(k)^2 \tilde R(k)}= {\delta_{ij}\over d}\int d{\bf k}\  k^2{\tilde K^2(k)\over \tilde\Delta(k)^2 \tilde R(k)}.
\end{equation}
We may thus write the effective Langevin equation for ${\bf x}(t) $ as
\begin{equation}
\left(1 + {h^2\kappa\over \kappa_\phi d(2\pi)^d}\int d{\bf k}\  k^2{\tilde K^2(k)\over \tilde\Delta(k)^2 \tilde R(k)}\right){dx_i(t)\over dt}
=  h\kappa\sqrt{{1\over \kappa_\phi}}\nabla_iK(R\Delta)^{-1}\xi({\bf x}(t),t) +\sqrt{\kappa}\eta_i(t)
\label{dec}
\end{equation}

Let us remark here that if we had {\em not} taken into account the effect of the particle position on the field and had simply considered the effect of the field on the particle we would have arrived at the 
effective diffusion equation
\begin{equation}
 {dx_i(t)\over dt}
=  h\kappa\sqrt{{1\over \kappa_\phi}}\nabla_iK(R\Delta)^{-1}\xi({\bf x}(t),t) +\sqrt{\kappa}\eta_i(t)
\label{deu}
\end{equation}
The effective diffusion constant for the process of Eq. (\ref{dec}) $\kappa_e$ is simply related to that
of Eq. (\ref{deu}), $\kappa^*$ via
\begin{equation}
\kappa_e = {\kappa^*\over \left[1 + {h^2\kappa\over \kappa_\phi d(2\pi)^d}\int d{\bf k}\  k^2{\tilde K^2(k)\over \tilde\Delta(k)^2 \tilde R(k)}\right]^2}
\end{equation}
If we write Eq. (\ref{deu}) as
\begin{equation}
{dx_i(t)\over dt}= \zeta_i({\bf x}(t),t),
\end{equation}
we see that the correlation function for the noise is given by
\begin{equation}
\langle \zeta_i({\bf x}(t),t)\zeta_j({\bf x}(t'),t')\rangle = 
2T\delta(t-t')\delta_{ij}\kappa\left[1 + {h^2\kappa\over \kappa_\phi d(2\pi)^d}\int d{\bf k}\  k^2{\tilde K^2(k)\over \tilde\Delta(k)^2 \tilde R(k)}\right],
\end{equation}
which immediately yields
\begin{equation}
\kappa^* = \kappa\left[1 + {h^2\kappa\over \kappa_\phi d(2\pi)^d}\int d{\bf k}\  k^2{\tilde K^2(k)\over \tilde\Delta(k)^2 \tilde R(k)}\right] \label{link}
\end{equation}
and 
\begin{equation}
{\kappa_e \over \kappa}= {1 \over \left[1 + {h^2\kappa\over \kappa_\phi d(2\pi)^d}\int d{\bf k}\  k^2{\tilde K^2(k)\over \tilde\Delta(k)^2 \tilde R(k)}\right]}.\label{ke}
\end{equation}
We thus see that if the particle is not coupled to the field that the the diffusion constant ($\kappa^*$) 
is increased, however when the coupling is taken into account the diffusion constant is reduced. More over to first  order in $\kappa/\kappa_\phi$ we find that the change in the two different diffusion constants
with respect to their bare values is the same in magnitude but of opposite sign. It is interesting to note that there is no  temperature dependence on the renormalization of  $\kappa_e$ due to the interaction
with the field. This means that $D_e$ retains a simple linear dependence on $T$ within the adiabatic approximation.

Note that the integrals occurring in the above can be ultra-violet (UV) divergent. This may the case for certain local operators $K$, for example $K({\bf x}-{\bf x}) = \delta({\bf x}-{\bf x}')$. When numerically
simulating the system in this case one can just introduce an ultra-violet cut-off in the simulation, {\em i.e.}
a maximal Fourier mode. The result given here shows that a naive application of the Stokes Einstein relationship works for the effective diffusion constant $D_e$. Define by $D$ the diffusion constant without a coupling to the field. For a particle moving at constant velocity the frictional force, opposing the  motion, is given by
\begin{equation}
f_0= \lambda_0 v
\end{equation}
However, the friction is related to the mobility via
\begin{equation}
\mu_0 = f_0v
\end{equation}.

Stokes Einstein tells us that, when it is valid (see later discussion),
\begin{equation}
D=\mu_0 T,
\end{equation}
and using the fact that $D=T\kappa$ we have that $\lambda_0 = 1/\kappa$

Now in presence of the coupling one can compute the average frictional force $f_\phi$ due to the fluctuation field \cite{dem2010a,dem2010b} and one finds that
\begin{equation}
f_\phi = \lambda_\phi v
\end{equation}
where $\lambda_\phi$ is given by
\begin{equation}
\lambda_\phi = {h^2 \over \kappa_\phi d(2\pi)^d}\int d{\bf k}\  k^2{\tilde K^2(k)\over \tilde\Delta(k)^2 \tilde R(k)}
\end{equation}
The total frictional force is thus given by
\begin{equation}
f=f_0 +f_\phi = (\lambda_0+\lambda_\phi)v,
\end{equation}
This gives via the Stokes-Einstein relation
\begin{eqnarray}
D_e = \mu_e T &=& {T \over \lambda_0+\lambda_\phi} \nonumber \\
&=& {D \over \left[1 + {h^2 D\beta\over \kappa_\phi d(2\pi)^d}\int d{\bf k}\  k^2{\tilde K^2(k)\over \tilde\Delta(k)^2 \tilde R(k)}\right]},\label{dab}
\end{eqnarray}
where $\beta=1/T$, which is equivalent to the result Eq. ({\ref{ke}). We also note that all the terms in
the integrand of Eq. ({\ref{ke}) are positive and thus we have within the adiabatic approximation that
$D_e<D$. The application of the Stokes-Einstein relation that we have just made is clearly not exact.
To compute the diffusion constant using the Stokes-Einstein relation one must compute the average 
value of the velocity $v$ at constant applied force \cite{revq}. This is a much harder problem than 
computing the average force at constant velocity.  In a previous paper \cite{dem2010b} we argued that the Stokes Einstein relation as applied above should be valid when the fluctuations of the force are small and thus the force is near to constant in the statistical sense. The fact that the average force is large means that the friction is large and thus the diffusion constant is small. Here we see that the 
adiabatic limit reproduces the approximate application of the Stokes-Einstein relation given above. 
This result can be explained by examining the expression given for the autocorrelation function of the force fluctuations given in \cite{dem2010a}, here one sees in the adiabatic limit that fluctuations of the force become uncorrelated in time and their amplitude becomes small.

\subsection{Examples}

We will now consider a number of special cases of our principle result Eq. (\ref{ke}) which we will write as
\begin{equation}
\kappa_e = {\kappa \over \left[1 + {h^2S_d\kappa Q\over \kappa_\phi d(2\pi)^d}\right]}.\label{ke2}
\end{equation}
with
\begin{equation}
Q = \int dk\  k^{d+1}{\tilde K^2(k)\over \tilde\Delta(k)^2 \tilde R(k)},
\end{equation}
and where $S_d$ is the area of a sphere of radius $1$ in $d$ dimensions. 

Considering now a localized magnetic field diffusing in a ferromagnet within the Gaussian
approximation, we take
\begin{equation}
{\tilde \Delta}(k) = k^2 + m^2
\end{equation}
we assume a magnetic field with a localized Gaussian profile and thus
\begin{equation}
{\tilde K}(k)= \exp\left(-{k^2 a^2\over 2}\right).
\end{equation} 
With this we find
\begin{equation}
Q = m^{d-2}\int_0^\infty dq {q^{d+1}\over (q^2+1)^2}\exp(-q^2m^2 a^2) \label{qa}
\end{equation}
for model A dynamics
and 
\begin{equation}
Q = m^{d-4}\int_0^\infty dq {q^{d-1}\over (q^2+1)^2}\exp(-q^2m^2 a^2)\label{qb}
\end{equation}
for model B dynamics. 

In certain cases one can take the limit $a\to 0$ in the above and thus obtain results that are 
only weakly dependent on the cut-off. However this cases depend strongly on the mass of the field theory and the result can be divergent when the theory is critical, i.e. when $m=0$. These cases are the following 
\begin{itemize}
\item{\em Model A: $d\leq 1$}
\begin{equation}
Q={\pi\over 4 m}, \ \ \  d=1.
\end{equation}

\item{ \em Model B: $d\leq 3$}
\begin{eqnarray}
Q&=& {\pi\over 4m^3}, \ \ \  d=1\nonumber \\
&=& {1\over 2 m^2}, \ \ \  d=2\nonumber \\
&=& {\pi\over 4m}, \ \ \ \;  d=3.
\end{eqnarray}
\end{itemize}

We thus see that as the mass of the scalar field is decreased or its correlation length $\xi=1/m$ increases  the diffusion constant of the active tracer particle is decreased. In the limit where $h\gg1$ the magnetic tracer will be surrounded by a polarized region where the field $\phi$
has the same sign as the tracer field. The size of the polarized region will be of order $\xi$ and 
the modification to the diffusion constant above presumably reflects the effective mobility of this
polarization cloud. It is also interesting to note the different dependence on $m$ between model
A in B in one dimension.

Cases where the results have a strong dependence on the cut-off have been extensively discussed in \cite{dem2010b}.

\section{Weak coupling limit}

In this section we use a formally Kubo-like expression for the effective diffusion constant, 
the expression can formally be computed to $O(h^2)$ in the particle-field coupling parameter, thus 
giving an expression for $D_e$ which is exact to this order. To obtain the Kubo formula we integrate Eq. ({\ref{xd}) to obtain
\begin{equation}
{\bf x}(t)-{\bf x}(0) = h\kappa\int_0^t \nabla K\phi({\bf x}(s)) ds + \sqrt{2T\kappa}{\bf B}(t) \label{inte}
\end{equation}
where ${\bf B}_t$ is a standard $d$ dimensional Brownian motion with
\begin{equation}
\langle B_i(t)B_j(s)\rangle =\delta_{ij}\min\{t,s\}.
\end{equation}
In the above we assume that at $t=0$ the system is in equilibrium (i.e. we assume that the
volume of the system is finite and ${\bf x}_0$ and the initial field configuration $\phi$ are chosen
from the equilibrium distribution (see \cite{revq} for more details). Now subtracting the first term of the
right hand side of Eq. (\ref{inte}) from both sides, squaring the resulting equation and taking the average
yields
\begin{equation}
\langle ({\bf x}(t)-{\bf x}(0))^2\rangle  + 
h^2\kappa^2\left\langle\left(\int_0^\infty \nabla K\phi({\bf x}(s)) ds\right)^2\right\rangle
=2Td\kappa t,
\end{equation}
where the cross term on the right had side is zero due to the Onsager relations \cite{revq}. 
We can thus define a time dependent diffusion constant via
\begin{equation}
\langle ({\bf x}(t)-{\bf x}(0))^2\rangle=2dD_e(t)t= 2dT\kappa_e(t)  
= 2Td\kappa t - h^2\left\langle\left(\kappa\int_0^t \nabla K\phi({\bf x}(s)) ds\right)^2\right\rangle
\end{equation}
where the late time limit of these two quantities are the effective values {\em}
$\lim_{t\to\infty} D_e(t),\ \kappa_e(t)=D_e,\ \kappa_e$.  We may therefore write
\begin{equation}
{D_e\over D} = 1 - {h^2\beta^2 D\over 2d}\lim_{t\to\infty}\left\langle{1\over t}\left(\int_0^t \nabla K\phi({\bf x}(s)) ds\right)^2\right\rangle. \label{kubo}
\end{equation}
>From this exact formula we see that the value of the diffusion constant is reduced by the interaction with
the field. This expression may now be evaluated to $O(h^2)$ by replacing ${\bf x}(t)$ by
the pure Brownian motion $\sqrt{2D}{\bf B}_t$ of the particle without interaction of the field
and using the correlation function for the free field without interaction with the particle which 
can be written as
\begin{equation}
\langle \phi_0({\bf x},t)\phi_0({\bf y},s)\rangle = T\int {d{\bf k}\over (2\pi)^d}
\tilde \Delta^{-1}(k)\exp\left(-\kappa_\phi|t-s|\tilde R(k)\tilde \Delta(k)\right)\exp\left(i{\bf k}\cdot({\bf x}-{\bf y})\right).
\end{equation}
After a straight forward computation  using the fact that ${\bf B}(t)$ and $\phi_0$ are independent we 
obtain
\begin{equation}
{D_e\over D} = 1 - {h^2\beta D\over d}  \int {d{\bf k}\over (2\pi)^d}{k^2\tilde K^2 (k)\over \tilde \Delta(k)
\left(\kappa_\phi \tilde R(k)\tilde \Delta(k) + D k^2\right)}+O(h^4).\label{ddp}
\end{equation}
Note that in the adiabatic limit this result is clearly equivalent to Eq. (\ref{dab}) to $O(h^2)$.
In terms of the variables $\kappa$ Eq. (\ref{ddp}) reads
\begin{equation}
{\kappa_e\over \kappa} = 1 - {h^2\kappa\over d}  \int {d{\bf k}\over (2\pi)^d}{k^2 \tilde K^2(k)\over \tilde \Delta(k)
\left(\kappa_\phi \tilde R(k)\tilde \Delta(k) + T\kappa k^2\right)}+O(h^4),\label{ddp}
\end{equation}
note therefore in contrast to the purely adiabatic result Eq. (\ref{ke})  that there is a 
temperature dependent renormalization of  $\kappa_e$. Another interesting thing to note is that
in this perturbative result we can recover the case where the field is frozen, i.e. the other extreme
to the adiabatic limit where $\kappa_\phi=0$. Here the field is quenched and has correlation function
$T\Delta^{-1}$. This quenched result agrees with the first order perturbation result for quenched 
random fields \cite{revq}.

\section{A Toy Model}

\begin{figure}
\begin{center}
\resizebox{0.8\hsize}{!}{\includegraphics[angle=0]{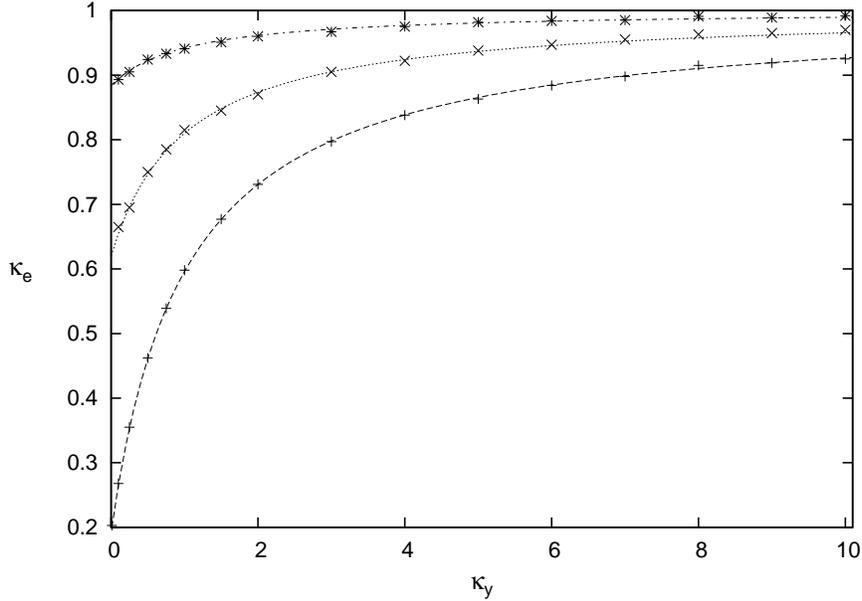}}
\end{center}
\caption{Effective diffusion coefficient versus field evolution velocity for the model of Eqs. (\ref{toy1}) and (\ref{toy2}) in one dimension for $V(x)=\cos(x)$ and temperatures $T=0.5$ (+), 1 ($\times$) and 2 (*). The continuous lines correspond to the analytical result.} 
\label{diffcos}
\end{figure}

In the general class of models studied above we have been able to obtain partial results on the 
effective diffusion constant on an active tracer in two distinct  limits, the adiabatic limit and the 
weak coupling limit. Here we present a simple toy model whose behavior can be thoroughly
analyzed. We consider a tracer particle ${\bf x}$ coupled to another diffusion process ${\bf y}$ 
via the Hamiltonian defined by
\begin{equation}
H({\bf x},{\bf y}) = V({\bf x}-{\bf y}),
\end{equation}
where $V$ is a function such that there exists a vector (or set of vectors) $\bf a$ such that, $V({\bf x})=V({\bf x}+{\bf a})$ in the algebraic or statistical sense.
For example we could take $V({\bf x})$ to be  a periodic function or one that is statistically
invariant by translation. The coupled diffusion
equations for $\bf x$ and $\bf y$ are given by 
\begin{eqnarray}
{\partial x_i\over \partial t}&=& -\kappa_x{\partial V({\bf x}-{\bf y})\over \partial x_i}+\sqrt{\kappa_x}\label{toy1}\eta_{x_i} \\
{\partial y_i\over \partial t}&=& \kappa_y{\partial V({\bf x}-{\bf y})\over \partial x_i}+\sqrt{\kappa_y}\eta_{y_i},\label{toy2}
\end{eqnarray}
where the noise variables above are white noise at temperature $T$ as defined earlier.
In the case where the variable $\bf y$ is frozen (or equivalently $\kappa_y=0$), as the function
$V$ is translationally invariant and if it is bounded, we expect the process $\bf x$  will 
have an effective quenched diffusion constant defined by
\begin{equation}
\langle {\bf x}_t^2\rangle \sim 2d D_e^{(q)}t  =2d T\kappa_e^{(q)}t,
\end{equation}
and which is independent of the choice of ${\bf y}$ by the translational invariance of $V$. 
We note that it is possible to compute $D_e$ exactly in a number of special cases \cite{revq}. In order to
see the effect of the dynamics of $\bf y$ on the process $\bf x$ we define the new variables
\begin{eqnarray}
{\bf u}&=& {\bf x}-{\bf y} \\
{\bf v} &=& \kappa_y {\bf x} +\kappa_x{\bf y},
\end{eqnarray}
 and it is then easy to see that these new dynamical variable obey
 \begin{eqnarray}
 {\partial u_i\over \partial t}&=& -(\kappa_x+\kappa_y){\partial V({\bf u})\over \partial u_i}+\sqrt{\kappa_x}\eta_{x_i} -\sqrt{\kappa_y}\eta_{y_i} \\
 {\partial v_i\over \partial t}&=&\kappa_y \sqrt{\kappa_x}\eta_{x_i} +\kappa_x\sqrt{\kappa_y}\eta_{y_i}.
 \end{eqnarray} 
 Furthermore one can easily see that the noise in these two equations are independent and thus
 the processes $\bf u$ and $\bf v$ are independent. The process $\bf u$ is simply a time rescaled 
 form of the quenched problem and $\bf v$ is a free Brownian motion. The mean squared displacement
 of the two processes can thus be computed easily and are given by
 \begin{eqnarray}
\langle {\bf u}_t^2\rangle &\sim& {2 d T\kappa_e^{(q)} }{\kappa_x +\kappa_y \over \kappa_x}t \\
\langle {\bf v}_t^2\rangle &\sim& 2dT \kappa_x\kappa_y(\kappa_x+\kappa_y)t.
\end{eqnarray} 
Finally using the independence of ${\bf u}$ and ${\bf v}$ we find that 
\begin{equation}
\langle {\bf x}_t^2\rangle \sim  2dT\kappa_x\left({\kappa_e^{(q)} +\kappa_y\over \kappa_x+\kappa_y}\right)t,
\end{equation}
which gives the effective diffusion constant of ${\bf x}$ as
\begin{equation}
{D_e\over T} =\kappa_e =  \kappa_x\left({\kappa_e^{(q)} +\kappa_y\over \kappa_x+\kappa_y}\right).
\label{toyres}
\end{equation}
We see that in the quenched limit $\kappa_y=0$ we obtain that $\kappa_e = \kappa_e^{(q)}$ as we should. In addition it is clear that $\kappa_e$ is an increasing function of $\kappa_y$, the quenched case being a lower bound for the effective diffusion constant. Another interesting fact about the 
expression Eq. (\ref{toyres}) is that there are a number of models where the quenched diffusion
constant vanishes signaling the transition from a regime of normal diffusion to one of subdiffusion
\cite{revq,tou2009,tou2007,tou2008}, however we see from Eq. (\ref{toyres}) that when $\kappa_y\neq
0$ then the vanishing of $\kappa_e^{(q)}$ does not cause the diffusion constant to vanish as the result
has an additive property. Indeed if $\kappa_e^{(q)}=0$ then we find 
\begin{equation}
\kappa_e =  {\kappa_x\kappa_y\over \kappa_x+\kappa_y}
\end{equation}
throughout the parameter region where the quenched problem shows subdiffusion.

The  quenched diffusion constant $\kappa_e^{(q)}$ can be computed exactly in a limited number
of cases \cite{revq}, notably in one dimension where it is given by
\begin{equation}
{\kappa_e^{(q)}\over \kappa_x} = {1\over \langle \exp(-\beta V)\rangle\langle \exp(\beta V)\rangle}
\label{1dqd}
\end{equation}
where
\begin{equation}
\langle \exp(\pm\beta V)\rangle = \lim_{L\to \infty}{1\over L}\int_0^L\exp(\pm\beta V(x))dx
\end{equation}
which exists for translationally invariant potentials. In addition when the field $V$ is equivalent 
to $-V$ (either functionally or statistically) a exact result in two dimensions gives
\begin{equation}
{\kappa_e^{(q)}\over \kappa_x} = {1\over \langle \exp(\beta V)\rangle}.
\end{equation}

As a test of this result in one dimension we took the potential $V(x) =\cos(x)$ and Eqs. (\ref{toyres}) and (\ref{1dqd}) then give
\begin{equation}
\kappa_e =  \kappa_x\left({{\kappa_x\over I^2_0(\beta)} +\kappa_y\over \kappa_x+\kappa_y}\right),
\label{ftoy}
\end{equation}
where $I_0(z)$ denotes the modified Bessel function. Numerical simulations were carried out for  $\kappa_x=1$ and  $\kappa_y$ varying between $0.1$ and $10$ and at temperatures $T=0.5$, 1 and 2.  The results given Figs. (\ref{diffcos}) show an excellent agreement with the analytical result Eq. (\ref{ftoy}).

\section{Discussion}
We have examined the diffusion of an active tracer particle coupled to a fluctuating field. Most
previous studies have been carried out on passive tracers in time dependent or quenched fields.
The action of the tracer on the field means that diffusion is always slowed down with respect to
the non-interacting case, this fact is explicit in Eq. (\ref{kubo}). This result may seem odd from a 
physical point of view as one would naively expect that the fluctuating field would help the 
particle to diffuse. Indeed we have seen that in the adiabatic approximation a passive diffuser
diffuses more quickly when driven by the field. However when the effect of the tracer on the field
is taken into account the effect of the extra noise from the fluctuating field is eliminated from a 
additional drag due to the action of the tracer on the field. Although it is in a limiting case the adiabatic 
calculation carried out here shows how these two effects compete and lead to slowing down of the diffusion with respect to the free case. As an example we analysed the diffusion constant of 
magnetic fields diffusing in Gaussian ferromagnets, here we found that the diffusion rate 
decreases as the correlation length of the field increases. This result is presumably linked to the fact
that the field is trapped in domains where the field has the same sign and also its presence
leads to the formation of these domains about it. As the correlation length increases the size of
the domains containing the field increases and the diffusion constant of the field effectively becomes
that of it surrounding domain. It seems physically reasonable that the diffusion constant of the surrounding domain becomes smaller as its size increases. 

We have also analysed the active diffusion problem in the weak coupling limit where the existence
of an underlying Gibbs measure allows us to write down a Kubo-type formula for the diffusion
constant. As well as rigorously establishing that diffusion is slowed down with respect to the free case,
this formula can be used to give a first order expression for the modified diffusion constant. 

Finally we have analysed a toy model for a particle interacting with a scalar field in the simple case where this scalar field is another diffusing particle. In this case the effective diffusion constant can
be formally computed in terms of the effective diffusion constant for a particle diffusing in 
a quenched random potential. The effect of the dynamics of the second diffuser can be thoroughly
understood and as its bare diffusion constant is increased so is that of the tracer.  

Clearly there are still a large number of questions about diffusion of active scalars in fluctuating fields, 
for the precise problem examined here the whole regime beyond the perturbative and adiabatic regimes
studied here remain open. It would also be interesting to understand in more detail the cross over from
the active to passive cases and understand under what circumstances field  fluctuations increase/decrease the tracer diffusion constant.

\noindent {\bf Acknowledgments:} 
DSD acknowledges support from the Institut Universitaire de France.

\end{document}